# Multicarrier Modulation-Based Digital Radio-over-Fibre System Achieving Unequal Bit Protection with Over 10 dB SNR Gain

Yicheng Xu[(1)], Yixiao Zhu[(1)], Xiaobo Zeng[(1)], Mengfan Fu[(1)], Hexun Jiang[(1)], Lilin Yi[(1)], Weisheng Hu[(1, 2)], Qunbi Zhuge*[(1, 2)]

[(1)] State Key Laboratory of Advanced Optical Communication Systems and Networks, Department of Electronic Engineering, Shanghai Jiao Tong University, Shanghai, China, *qunbi.zhuge@sjtu.edu.cn
[(2)] Peng Cheng Laboratory, Shenzhen, China

**Abstract** *We propose a multicarrier modulation-based digital radio-over-fibre system achieving unequal bit protection by bit and power allocation for subcarriers. A theoretical SNR gain of 16.1 dB is obtained in the AWGN channel and the simulation results show a 13.5 dB gain in the bandwidth-limited case.* ©2023 The Author(s)

## Introduction

As a significant segment in the centralized radio access network (C-RAN), fibre-based fronthaul links provide the connection between baseband units and remote radio units [1, 2]. Digital radio-over-fibre (DRoF) is a typical solution for fronthaul, which digitizes the wireless signals into bit sequences and transmits them through a fibre channel [3]. Recently, in order to satisfy the increasing capacity demands for fronthaul, multi-level modulation formats such as 4-level pulse amplitude modulation (PAM-4), have been taken into account [4, 5]. However, multi-level modulation formats are more vulnerable to link noise due to the reduced Euclidean distance, which means more challenging realization of error-free transmission.

On the other hand, a notable feature of DRoF is that several bits in the quantized signals hold greater significance than other bits, and errors in these bits can lead to more severe performance degradation of wireless signals. Based on the premise, several approaches have been proposed that offer unequal bit protection to more significant bits (MSBs). In [6], unequally-spaced PAM-4 with bits interleaving is adopted, which maps the MSBs into the more reliable bit of PAM-4 symbols. A time-domain hybrid PAM-2/4 scheme [7] has also been proposed to transmit the MSBs with PAM-2 format. These approaches are mainly employed in a single carrier scheme, while a multicarrier scheme [8] offers more dimensions for flexible adjustment.

In this paper, we propose a multicarrier modulation-based digital radio-over-fibre (MCM-DRoF) system, which provides unequal protection for MSBs by bit and power allocation for subcarriers. MCM has advantages as follows: 1) flexible adjustment of digital subcarrier (DSC) performance. 2) inner DSCs are less affected by filtering effects than outer ones. These advantages provide natural ways for unequal protection of MSBs. A 16.1 dB signal-to-noise ratio (SNR) gain of the MCM-DRoF over single carrier DRoF (SC-DRoF) is illustrated in theory. Furthermore, simulation results show that the MCM-DRoF system with only bit allocation achieves a SNR gain of 4.9 dB. With both bit and power allocation, the SNR gain reaches 13.5 dB.

## Principle

Fig. 1 shows the schematic diagram of our proposed MCM-DRoF system. The in-phase (I) and quadrature (Q) parts of the complex wireless signals, namely I/Q samples, are first separated and then uniformly digitized into bit sequences. In the following instructions, we take the 12-bit quantization as an example. Each quantized I/Q sample is mapped into a 12-bit binary codeword $b_{11}b_{10} \cdots b_1 b_0$, in which the importance of bits descends from high-index bits to low-index bits. The foremost bit $b_{11}$ represents the sign of I/Q samples, and other bits represent the amplitude of I/Q samples. Note that a scrambler is used to generate bits with equal probabilities of 0 and 1. Then these bit sequences are allocated to different subcarriers. In this example, we use three subcarriers and the modulation format is 16-ary quadrature amplitude modulation (16-QAM) for all subcarriers. The first subcarrier consists of binary codeword $b_{11}b_{10}b_9b_8$, which is the most important one, and we denote it as high-priority digital subcarrier (HP-DSC). $b_7b_6b_5b_4$ and $b_3b_2b_1b_0$ in the codeword are allocated to other two subcarriers denoted as medium-priority digital subcarrier (MP-DSC) and low-priority digital subcarrier (LP-DSC), respectively. Note that bit interleaving within each subcarrier is considered before 16-QAM modulation.

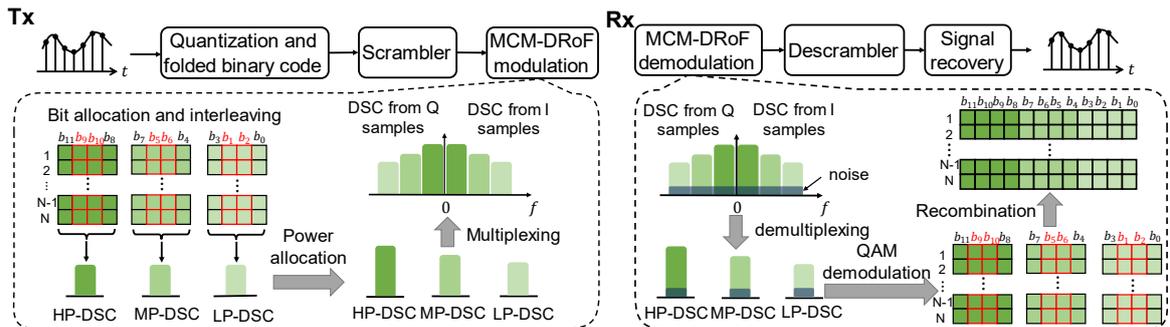

**Fig. 1:** Schematic diagram of the proposed MCM-DRoF system. HP/MP/LP: high/medium/low-priority. DSC: digital subcarrier.

Afterwards, power allocation for subcarriers is accomplished. We define that the power ratio of three subcarriers is $p_1:p_2:p_3$, and $p_1$ is set to 1 as the reference. The optimization criterion of power factors $(p_2, p_3)$ is further explained in the theoretical analysis. In the multiplexing process, HP-DSC, MP-DSC, and LP-DSC are placed at the innermost, middle, and outmost positions of the spectrum, respectively. This structure is helpful to reduce the penalty of low-pass filtering effects caused by the transceiver. For coherent transmission, subcarriers representing I and Q samples are placed in the positive and negative parts of the frequency domain, respectively, forming a symmetrical spectrum with six subcarriers. At the receiver, the reverse of the above operations is applied, as shown in Fig. 1.

**Theoretical analysis**
In the DRoF system, the SNR of recovered I/Q samples can be calculated as

$$SNR_r = \frac{P_s}{N_q + N_e} \quad (1)$$

where $P_s$ is the signal power. $N_q$ and $N_e$ are the quantization noise and bit-error-induced noise, respectively. $N_q$ is equal to $\frac{\Delta^2}{12}$ and $\Delta$ is the quantization interval [9]. We focus on the calculation of bit-error-induced noise. Error bits in the received binary codeword will lead to the de-mapping error of I/Q samples, which is equivalent to an additive noise to the original I/Q samples. For the power of bit-error-induced noise, it is related to the error bit position in a binary codeword and its bit error ratio (BER). Tab. 1 lists the noise powers of each bit in a 12-bit folded binary codeword and we refer to it as weighting factors of each bit. It can be observed that the weighting factors descend from high-index bits to low-index bits, which means greater importance of high-index bits. Hence, the power of bit-error-induced noise for $L$-bit quantization can be calculated as

$$N_e = \sum_{m=0}^{L-1} P_{e_m} W_m \quad (2)$$

where $P_{e_m}$ and $W_m$ are the BER and weighting factor of the $m^{th}$ bit in the codeword, respectively. The final step is to derive $P_{e_m}$ from SNRs of subcarriers. Note that an additive white Gaussian noise (AWGN) channel is used in our theoretical analysis. Therefore, under the average power constraint, the SNRs of subcarriers after power allocation can be calculated as:

$$SNR_{sub_i} = \frac{n_{sub} \cdot p_i}{\sum_{i=1}^{n_{sub}} p_i} \cdot SNR_{channel} \quad (3)$$

where $SNR_{sub_i}$ and $SNR_{channel}$ are the SNRs of the $i^{th}$ subcarrier and the AWGN channel, respectively. $n_{sub}$ is the number of subcarriers. $p_i$ is the power factor of the $i^{th}$ subcarrier as mentioned before. Since 16-QAM is adopted for each subcarrier, $P_{e_m}$ can be calculated from $SNR_{sub_i}$. Based on the above equations, the optimization target of power factors $(p_2, p_3)$ is to maximize $SNR_r$ or namely minimize $N_e$.

**Tab. 1:** Weighting factors (dBm) of different bits in a codeword when the peak value of signals is 1.

| $W_{11}$ | $W_{10}$ | $W_9$ | $W_8$ | $W_7$ | $W_6$ |
|---|---|---|---|---|---|
| 24.1 | 24.0 | 18.0 | 11.9 | 5.9 | -0.1 |
| $W_5$ | $W_4$ | $W_3$ | $W_2$ | $W_1$ | $W_0$ |
| -6.1 | -12.1 | -18.2 | -24.2 | -30.2 | -36.2 |

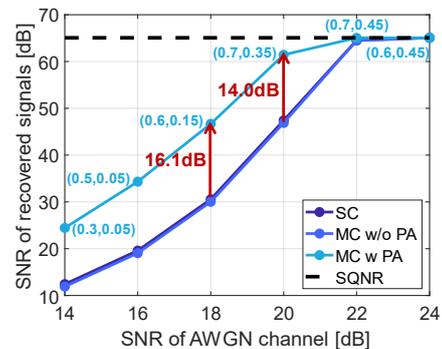

**Fig. 2:** Theoretical results with the AWGN channel.
SC: single carrier. MC: multicarrier. PA: power allocation.
SQNR: signal to quantization noise ratio.

Fig. 2 shows the theoretical results of MCM-DRoF compared with single carrier based 16-QAM DRoF. The optimized power factors $(p_2, p_3)$ for each channel SNR are marked in the figure. As the channel SNR decreases, the performance of SC-DRoF degrades rapidly due to the rising BER, while MCM-DRoF exhibits a relatively small decrease, thanks to the unequal bit protection provided by power allocation. When the channel SNR is 18 dB, a gain of 16.1 dB is achieved, which improves the highest modulation format of wireless signals from 256-QAM to 16384-QAM [10]. The comparison of BERs with and without power allocation at 18 dB channel SNR is listed in Tab. 2, in which an obvious decrease in the BERs of MSBs is observed.

**Tab. 2:** BERs of the codeword w/o and w/ power allocation.

| | $b_{11}$&$b_{10}$ | $b_9$&$b_8$ | $b_7$&$b_6$ | $b_5$&$b_4$ | $b_3$&$b_2$ | $b_1$&$b_0$ |
|---|---|---|---|---|---|---|
| w/o PA | 9.5x10⁻⁵ | 1.9x10⁻⁴ | 9.5x10⁻⁵ | 1.9x10⁻⁴ | 9.5x10⁻⁵ | 1.9x10⁻⁴ |
| w/ PA | 8.3x10⁻⁷ | 1.7x10⁻⁶ | 7.9x10⁻⁵ | 1.6x10⁻⁴ | 1.7x10⁻² | 3.5x10⁻² |

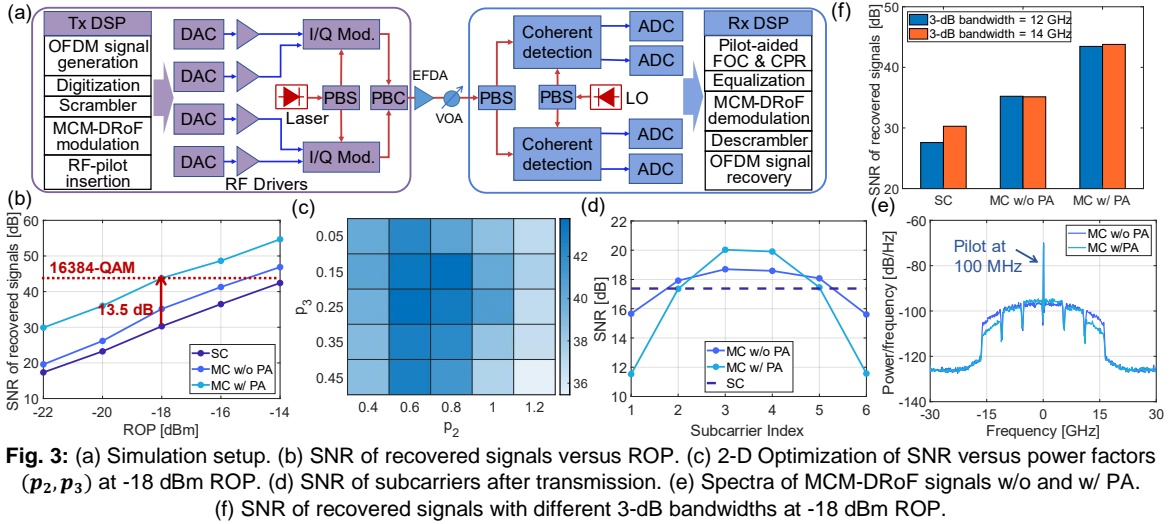

**Fig. 3:** (a) Simulation setup. (b) SNR of recovered signals versus ROP. (c) 2-D Optimization of SNR versus power factors ($p_2, p_3$) at -18 dBm ROP. (d) SNR of subcarriers after transmission. (e) Spectra of MCM-DRoF signals w/o and w/ PA. (f) SNR of recovered signals with different 3-dB bandwidths at -18 dBm ROP.

**Simulation setup and results**

The simulation setup based on the optical coherent system is shown in Fig. 3(a). At the transmitter, after MCM-DRoF modulation, a pilot is inserted in the frequency domain for frequency offset compensation (FOC) and carrier phase recovery (CPR). The roll-off factor is set to 0.1. After passing through digital-to-analog converters (DACs), a 5-order Bessel filter is deployed to emulate the limited bandwidth of the transmitter. Then the digital signal is modulated into the optical domain by a dual-polarization IQ modulator. The linewidth of the laser is 100 kHz. At the receiver, the optical signal is detected by a coherent receiver followed by four analog-to-digital converters (ADCs). The power spectral density of thermal noise in the photodiode is $14 \times 10^{-12}$ A/Hz$^{1/2}$. The optical power of the local oscillator (LO) is 12 dBm. Its linewidth is 100 kHz and a frequency offset of 80 MHz is considered. An identical Bessel filter is used to emulate the limited bandwidth of the receiver. After pilot-aided FOC and CPR, the demultiplexed subcarriers pass through the equalizers, followed by MCM-DRoF demodulation and wireless signal recovery.

The simulation results are shown in Fig. 3(b)-(f). We have estimated the performance of SC-DRoF, MCM-DRoF without power allocation (only with bit allocation), and MCM-DRoF with bit and power allocation. The total symbol rate of the digital signal is set to 30 Gbaud and the 3-dB bandwidth of Bessel filters is 14GHz. As shown in Fig. 3(b), different from results in the AWGN channel model, MCM-DRoF without power allocation still shows superior performance over SC-DRoF, because the filtering effects caused by the limited bandwidth of the transceiver mainly affects the LP-DSC in the MCM-DRoF. At -18 dBm received optical power (ROP), with 13.5 dB gain after power allocation, the SNR of recovered wireless signals is 43.8 dB, which can satisfy the transmission of 16384-QAM wireless signals [10]. The 2-D optimization of power factors ($p_2, p_3$) at -18 dBm ROP is shown in Fig. 3(c) as an example. The SNR of subcarriers and spectra of the MCM-DRoF signals are shown in Fig. 3(d) and (e).

We also emulate the influence of different bandwidth limitations at -18 dBm ROP. In Fig. 3(f), SC-DRoF is more vulnerable to filtering effects with an obvious SNR degradation, while MCM-DRoF shows negligible SNR difference thanks to its resistance to filtering effects.

**Conclusions**

In this paper, we propose an MCM-DRoF system achieving unequal bit protection by bit and power allocation for digital fronthaul. In the theoretical analysis, a 16.1 dB SNR gain is obtained in the AWGN channel. In the simulation, the SNR of recovered wireless signals achieves 13.5 dB gain in the bandwidth-limited case. All the results indicate that the MCM-DRoF system can be a promising candidate for future digital fronthaul.

**Acknowledgements**

This work was supported by National Key R&D Program of China (2022YFB2903500), National Natural Science Foundation of China (62175145) and Shanghai Pilot Program for Basic Research - Shanghai Jiao Tong University (21TQ1400213).